\documentclass[prd,superscriptaddress,twocolumn,preprintnumbers,showpacs]{revtex4}

\usepackage{amssymb,amsmath,graphicx,epsfig,color}

\def\etal{{\em et al.}}
\def\issue(#1,#2,#3){{\bf #1}, #2 (#3)} 

\def\APP(#1,#2,#3){{\rm Acta Phys.\ Polon.} \ \issue({\bf #1},#2,#3)}
\def\ANP(#1,#2,#3){{\rm Annals of Physics} \ \issue({\bf #1},#2,#3)}
\def\ARNPS(#1,#2,#3){{\rm Ann.\ Rev.\ Nucl.\ Part.\ Sci.} \ \issue({\bf #1},#2,#3)}
\def\CPC(#1,#2,#3){{\rm Comp.\ Phys.\ Comm.} \ \issue({\bf #1},#2,#3)}
\def\CIP(#1,#2,#3){{\rm Comput.\ Phys.} \ \issue({\bf #1},#2,#3)}
\def\EPJ(#1,#2,#3){{\rm Eur.\ Phys.\ J.} \ \issue({\bf #1},#2,#3)}
\def\EPJD(#1,#2,#3){Eur.\ Phys.\ J. Direct\ C \ \issue({\bf #1},#2,#3)}
\def\IJMP(#1,#2,#3){{\rm Int.\ J.\ Mod.\ Phys.} \ \issue({\bf #1},#2,#3)}
\def\JHEP(#1,#2,#3){{\rm J.\ High Energy Physics} \ \issue({\bf #1},#2,#3)}
\def\JP(#1,#2,#3){{ J.\ Phys.} \ \issue({\bf #1},#2,#3)}
\def\MPL(#1,#2,#3){{Mod.\ Phys.\ Lett.} \ \issue({\bf #1},#2,#3)}
\def\NP(#1,#2,#3){{Nucl.\ Phys.} \ \issue({\bf #1},#2,#3)}
\def\NIM(#1,#2,#3){{ Nucl.\ Instrum.\ Meth.} \ \issue({\bf #1},#2,#3)}
\def\PL(#1,#2,#3){{ Phys.\ Lett.} \ \issue({\bf #1},#2,#3)}
\def\PR(#1,#2,#3){{ Phys.\ Rev.} \ \issue({\bf #1},#2,#3)}
\def\PRL(#1,#2,#3){{ Phys.\ Rev.\ Lett.} \ \issue({\bf #1},#2,#3)}
\def\SJNP(#1,#2,#3){{ Sov.\ J. Nucl.\ Phys.} \ \issue({\bf #1},#2,#3)}
\def\ZP(#1,#2,#3){{Zeit.\ Phys.} \ \issue({\bf #1},#2,#3)}
\def\be {\begin{equation}}
\def\ee {\end{equation}}
\def\bea {\begin{eqnarray}}
\def\eea {\end{eqnarray}}

\def\bdbdbar {B_d-\overline{B}_d}
\def\bsbsbar {B_s-\overline{B}_s}

\def\dgs {\Delta\Gamma_s}

\def\dms {\Delta M_s}

\def\m12np{M_{12}^{\rm NP}}
\def\g12np{\Gamma_{12}^{\rm NP}}

\definecolor{purple}{rgb}{0.63,0.13,0.94}
\definecolor{red}{rgb}{1.0,0.0,0.0}
\definecolor{green}{rgb}{0.0,1.0,0.0}
\definecolor{blue}{rgb}{0.0,0.0,1.0}

\begin{document}

\title{Reconciling anomalous measurements in $B_s-\overline{B}_s$ mixing: \\
the role of CPT-conserving and CPT-violating new physics} 

\author{Amol Dighe}
\email{amol@theory.tifr.res.in}
\affiliation{Tata Institute of Fundamental Research, 
Homi Bhabha Road, Colaba, Mumbai 400005, India}

\author{Diptimoy Ghosh}
\email{diptimoyghosh@theory.tifr.res.in}
\affiliation{Tata Institute of Fundamental Research, 
Homi Bhabha Road, Colaba, Mumbai 400005, India}

\author{Anirban Kundu}
\email{anirban.kundu.cu@gmail.com}
\affiliation{University of Calcutta,
92, Acharya Prafulla Chandra Road, Kolkata 700 009, India. }

\author{Sunando Kumar Patra}
\email{sunandoraja@gmail.com}
\affiliation{University of Calcutta, 
92, Acharya Prafulla Chandra Road, Kolkata 700 009, India. }

\date{\today}

\begin{abstract} 
Recently observed anomalies in the $B_s \to J/\psi \phi$ decay and 
the like-sign dimuon asymmetry $A^b_{sl}$ hint at possible 
new physics (NP) in the $\bsbsbar$ mixing. 
We parameterize the NP with four model-independent quantities: 
the magnitudes and phases of the dispersive part $M_{12}$ and the 
absorptive part $\Gamma_{12}$ of the NP contribution to the
effective Hamiltonian.
We constrain these parameters using the four observables
$\Delta M_s$, $\Delta\Gamma_s$, the mixing phase $\beta_s^{J/\psi\phi}$, 
and $A^b_{sl}$. 
Our quantitative fit indicates that the NP should contribute a
significant dispersive as well as absorptive part.
In fact, models that do not contribute a new absorptive part
are disfavored at more than 99\% confidence level. 
We extend this formalism to include CPT violation, 
and show that CPT violation by itself, or even in presence of 
CPT-conserving new physics without an absorptive part, 
helps only marginally in the simultaneous resolution of these anomalies.
The NP absorptive contribution to $\bsbsbar$ mixing therefore seems 
to be essential, and would imply a large branching fraction for
channels like $B_s \to \tau^+ \tau^-$.
\end{abstract}


\preprint{TIFR/TH/11-21}

\maketitle


\section{Introduction}
\label{sec1}

The Cabibbo-Kobayashi-Maskawa (CKM) paradigm of quark mixing in the standard
model (SM) is yet to be accurately tested in the $\bsbsbar$ sector, and it is
quite possible that the NP can affect the $\bsbsbar$ system 
while keeping the $\bdbdbar$ system untouched.
Indeed, for most of the flavor-dependent NP models,
the couplings relevant for the second and third generations of SM fermions
are much less constrained than those for the first generation fermions,
allowing the NP to play a significant role in the $\bsbsbar$ mixing,
in principle.

Over the last few years, the Tevatron experiments CDF and D\O~, and to a 
smaller extent the B factories Belle and BaBar, have provided
a lot of data on the $B_s$ meson, most of which are consistent with
the SM. There are some measurements, though, which show a significant
deviation from the SM expectations, and hence point towards new physics (NP).
The major ones among these are the following.
(i) Measurements in the decay mode $B_s \to J/\psi \phi$ yield a large 
CP-violating phase $\beta_s^{J/\psi \phi}$ \cite{cdf-d0-note}.
In addition, though the difference 
$\Delta \Gamma_s$ between the decay widths of the mass eigenstates 
measured in this decay is consistent with the SM, it allows
$\Delta\Gamma_s$ values that are almost twice the SM prediction,
and also opposite in sign
\cite{hfag}.
(ii) The like-sign dimuon asymmetry $A^b_{sl}$ in the combined $B$ data 
at D\O~ \cite{d0dimuon} is almost $4\sigma$ away from the SM
expectation.

The resolutions of the above anomalies, separately or simultaneously,
have been discussed in the context of specific NP models:
a scalar leptoquark model \cite{dighe1,dighe2}, 
models with an extra flavor-changing neutral gauge boson $Z'$ 
or R-parity violating supersymmetry \cite{deshpande,alok-london},
two-Higgs doublet model \cite{dobrescu,bhaskar}, 
models with a fourth generation of fermions \cite{debchou,nandi}, 
supersymmetric grand unified models \cite{parry}, 
supersymmetric models with split sfermion generations \cite{endo} or
models with a very light spin-1 particle \cite{oh}. 
Possible four-fermion effective interactions that are 
consistent with the data have been analyzed by \cite{bauer} 
and the results are consistent with \cite{dighe2}. 
Similar studies, based on the minimal flavor violating (MFV) models
\cite{blum}, and the Randall-Sundrum model \cite{murugesh},
have been carried out. 

In this paper, we try to determine, in a model-independent way,
 which kind of NP would be able to
account for both the above anomalies simultaneously.
We take a somewhat different approach than the references 
cited above.
Rather than confining ourselves to specific models,
we assume that the NP responsible for the anomalies contributes
entirely through the $\bsbsbar$ mixing, and parameterize it
in a model independent manner through the effective Hamiltonian
for the $\bsbsbar$ mixing.
This effective Hamiltonian ${\cal H}$ is a $2\times 2$ matrix in
the flavor basis, and the relevant NP contribution appears
in its off-diagonal elements.
The NP can then be parameterized by using four parameters:
the magnitudes and phases of the dispersive part 
and the absorptive part 
of the NP contribution to ${\cal H}$.
A ``scatter-plot'' analysis that constrained these four new 
parameters using only $A^b_{sl}$ has been carried out in \cite{murugesh}.
We perform a $\chi^2$ fit to the $\bsbsbar$ mixing observables 
and obtain a quantitative measure for which kind of NP is
preferred by the data.
This would lead us to shortlist specific NP models that have the
desired properties, which can give testable predictions for
other experiments.
It is found that the NP needs to contribute to both the
dispersive as well as absorptive part of the Hamiltonian
in order to avoid any tension with the data.

We also extend our framework to include possible CPT violation
in the $\bsbsbar$ mixing, parameterized through the difference
in diagonal elements of ${\cal H}$. 
The motivation is to check if this can obviate the need for
an absorptive contribution from the NP.
Such an analysis to constrain CPT and Lorentz violating parameters
was carried out in \cite{kostel-bs}. 
However they have used only $A^b_{sl}$ and not $\beta_s^{J/\psi\phi}$ 
in their analysis, and their parameters are only indirectly
connected to the elements of ${\cal H}$.
We try to account for the two anomalies above with only CPT
violation as the source of NP, and with a combination of
CPT violation and the NP contribution to the off-diagonal
elements of ${\cal H}$. 
As we will show, nothing improves the
fit significantly from the SM unless there is a nonzero 
absorptive part in the $\bsbsbar$ mixing amplitude.

The paper is organized as follows. In Sec.~\ref{Hamiltonian},
we introduce our formalism for the four NP parameters.
In Sec.~\ref{measurements},
we summarize the experimental measurements and theoretical
predictions for the observables relevant for $\bsbsbar$ mixing.
In Sec.~\ref{fit}, we present the results of our fits, 
and their implications for NP models are discussed in Sec.~\ref{preferred}.
In Sec.~\ref{cptv-formalism}, we introduce the formalism for
introducing CPT violation and in Sec.~\ref{cptv-fit} we explore
the extent to which it can help resolving the anomalies.
Sec.~\ref{concl} summarizes our results and concludes.

\section{The effective Hamiltonian}
\label{Hamiltonian}

The evolution of a $\bsbsbar$ state can be described by the effective
Hamiltonian 
\be
{\cal H} = 
\begin{pmatrix} M_{11} & M_{12} \\ M_{12}^* & M_{22} \end{pmatrix}
- \frac{i}{2}
\begin{pmatrix} \Gamma_{11} & \Gamma_{12} \\ 
\Gamma_{12}^* & \Gamma_{22} \end{pmatrix}
\label{eff-H}
\ee
in the flavor basis, where $M_{ij}$ and $\Gamma_{ij}$ are its dispersive
and absorptive parts, respectively.
When CPT is conserved, $M_{11} = M_{22}$ and $\Gamma_{11}=\Gamma_{22}$.
The eigenstates of this Hamiltonian are $B_{sH}$ and $B_{sL}$,
with masses $M_{sH}$ and $M_{sL}$ respectively, and 
decay widths $\Gamma_{sH}$ and $\Gamma_{sL}$ respectively. 
The difference in the masses and decay widths can be written in terms
of the elements of the Hamiltonian as
\bea
\Delta M_s & \equiv & M_{sH}-M_{sL} \approx 2 |M_{12}| \; ,
\label{dms-def} \nonumber \\
\Delta\Gamma_s & \equiv & \Gamma_{sL} - \Gamma_{sH} \approx
2 |\Gamma_{12}| \cos[{\rm Arg}(-M_{12}/\Gamma_{12})] \, .
\label{dgs-def}
\eea
The above expressions are valid as long as $\Delta\Gamma_s \ll M_s$,
which is indeed the case here.

Since CPT is conserved, the effect of NP can be felt only through
the off-diagonal elements of ${\cal H}$. We separate the
SM and NP contributions to these terms via
\bea
M_{12} & = & M_{12}^{\rm SM} + M_{12}^{\rm NP} \; , \nonumber \\
\Gamma_{12} & = & \Gamma_{12}^{\rm SM} + \Gamma_{12}^{\rm NP} \; .
\label{sm-np}
\eea 
The NP can then be completely parameterized in terms of four real
numbers: $|\m12np|, \, {\rm Arg}(\m12np), \, |\g12np|$ and ${\rm Arg}(\g12np)$.
We take the phases ${\rm Arg}(\m12np)$ and ${\rm Arg}(\g12np)$ to
lie in the range 0-2$\pi$.

In a large class of models, including the Minimal Flavor Violation
(MFV) models,
the NP contribution has no absorptive part, i.e. 
$\Gamma_{12} = \Gamma_{12}^{\rm SM}$.
This is true for a lot of non-MFV models too.
This occurs when NP does not give rise to any new intermediate
light states to which $B_s$ or $\overline{B}_s$ can decay.
For such models, Eq.~(\ref{dgs-def}) implies that 
$\Delta\Gamma_s \lesssim \Delta\Gamma_s({\rm SM}) \approx 2 |\Gamma_{12}^{
\rm SM}|$,
i.e. the value of $\Delta \Gamma_s$ is always less than its 
SM prediction \cite{grossman}.
In such models, the NP is parameterized by only two parameters:
$|\m12np|$ and ${\rm Arg}(\m12np)$.
An analysis restricted to this class of models was performed in 
\cite{ligeti1}.

However there exists a complementary class of viable models where
the NP contributes to $\Gamma_{12}$ substantially.
These include models with leptoquarks, R-parity violating
supersymmetry, a light gauge boson, etc.
It has been pointed out in \cite{dighe1} that such a nonzero absorptive 
part that arises naturally in these class of models can enhance 
$\Delta\Gamma_s$ significantly above its SM value, contrary to 
the popular expectations based on \cite{grossman}. 
One notes that a new absorptive part in the mixing amplitude
necessarily means new final states that can be accessed by
both $B_s$ and $\overline{B}_s$. 
The data from the direct measurements of branching ratios 
is extremely restrictive \cite{bauer}, apart from that for a few 
final states like $B_s\to\tau^+\tau^-$ \cite{dighe2}.
As we shall see later in this paper, such models are favored
by the $\bsbsbar$ mixing data. 
The importance of $\tau^+\tau^-$ final states from $B_d$ and $B_s$ 
decays has also been pointed out in \cite{gln}.

\section{The measurements}
\label{measurements}

The $\bsbsbar$ oscillation and CP violation therein can 
be quantified by four observables, viz. the mass difference $\Delta M_s$, 
the decay width difference $\Delta\Gamma_s$, 
the CP-violating phase $\beta_s^{J/\psi\phi}$, and 
the semileptonic asymmetry $a^s_{\rm sl}$. 

The mass difference is measured to be
\be
\Delta M_s = (17.77\pm 0.10\pm 0.07)~{\rm ps}^{-1}\; ,
\label{dm-value}
\ee
which is consistent with the SM expectation \cite{lenz-nierste}
\be
\Delta M_s({\rm SM}) =  (17.3\pm 2.6)~{\rm ps}^{-1} \;.
\label{dm-sm}
\ee
However measurements in the $B_s \to J/\psi \phi$ decay mode
show a hint of some deviation from the SM.
The CP-violating phase $\beta_s^{J/\psi\phi}$ 
in this decay is
\be
\beta_s^{J/\psi\phi} = \frac{1}{2}
{\rm Arg} \left( -\frac{ (V_{cb}V_{cs}^\ast)^2}{M_{12}}
\right) \; , 
\label{betas-def}
\ee
whose average value measured at the Tevatron experiments
\cite{cdf-d0-note} is
\be
\beta_s^{J/\psi\phi} = (0.41^{+0.18}_{-0.15}) \cup (1.16^{+0.15}_{-0.18})\,.
\label{betas-value}
\ee
In the SM,
\be
\beta_s^{J/\psi\phi}({\rm SM}) = 
{\rm Arg} \left( -\frac{V_{cb}V_{cs}^\ast}{V_{tb}V_{ts}^\ast}
\right) \approx 0.019\pm 0.001\, .
\label{betas-sm}
\ee
Thus, the measured value of $\beta_s^{J/\psi \phi}$ is more than 
$2\sigma$ away from the SM expectation. On the other hand,
the difference in the decay widths of
the mass eigenstates $B_H$ and $B_L$ is measured to be \cite{cdf-d0-note}
\be
\Delta\Gamma_s = \pm (0.154^{+0.054}_{-0.070})
~{\rm ps}^{-1}\,,
\label{dgs-value}
\ee
while the SM expectation is \cite{lenz-nierste}
\be
\Delta\Gamma_s ({\rm SM}) =  (0.087\pm 0.021)~{\rm ps}^{-1} \,.
\label{dgs-sm}
\ee
The measurement is consistent with the SM expectation to $\sim 1 \sigma$,
however it allows for $\Delta\Gamma_s$ values that are almost twice the
SM prediction. 
Note that the sign of $\Delta\Gamma_s$ is undetermined experimentally
and this gives us more room to play with the NP parameters.

CDF has recently announced its new results, based
on 5.2 fb$^{-1}$ of data \cite{cdf-ichep}:
\bea
\vert \Delta\Gamma_s\vert &=& (0.075 \pm 0.035 \pm 0.010)~{\rm ps}^{-1}\,,
\nonumber\\
\beta_s^{J/\psi\phi} &=& (0.02 - 0.52) \cup (1.08 - 1.55)\,
\label{expdata-cdf}
\eea
to 68\% C.L..
While we note that the results are consistent with the SM, 
the final Tevatron averages are still awaited.
Therefore, we use the values in Eq.~(\ref{dgs-value}) in our analysis. 

The other anomalous measurement is the like-sign dimuon asymmetry.
Averaging the 9.0 fb$^{-1}$ data of D\O~ \cite{d0dimuon} and 1.6 fb$^{-1}$
data of CDF \cite{cdfdimuon}, and adding the errors in quadrature and treating
them as Gaussian, we get
\be
A^b_{\rm sl} =  -(7.41\pm 1.93)\times 10^{-3}\,,
\label{ab-value}
\ee
which differs by more than $3\sigma$ from its SM prediction
\be
A^b_{\rm sl} ({\rm SM})  = (-0.23^{+0.05}_{-0.06})\times 10^{-3} \,.
\label{ab-sm}
\ee
Note that for $A^b_{\rm sl}$, CDF has a poorer statistics than D\O~and 
therefore 
the average value is dominated by the D\O~data.

Even in the presence of new physics, the SM relationship holds:
\be
A^b_{\rm sl} = (0.506 \pm 0.043) a^d_{\rm sl} + (0.494\mp 0.043) a^s_{\rm sl}\,,
\label{ab-ad-as}
\ee
where $a^s_{\rm sl}$ and $a^d_{\rm sl}$ are the semileptonic asymmetries for the 
$\bsbsbar$  and the $\bdbdbar$ systems, respectively.
The former is related to the $\bsbsbar$ mixing observables through
\be
a^s_{\rm sl}  = \frac{\Delta\Gamma_s}{\Delta M_s} \tan\phi_s
\label{asl-def}
\ee
where $\phi_s \equiv {\rm Arg}(-M_{12}/\Gamma_{12})$.
The latter is defined analogously.
The coefficients in Eq.~(\ref{ab-ad-as}) are experimentally measured, 
and contain information about $\Delta M_{d(s)}$,
$\Delta\Gamma_{d(s)}$, and production fractions of $B_d$ and $B_s$ mesons.
Using $a^d_{\rm sl} = -(4.7\pm 4.6)\times 10^{-3}$ \cite{hfag},
this leads to
\be
a^s_{\rm sl} = - 0.010 \pm 0.006 \; , 
\label{asl-value}
\ee
which is about $1.7\sigma$ away from the SM prediction
\be
a^s_{\rm sl}({\rm SM}) = (2.06\pm 0.57)\times 10^{-5} \; .
\label{asl-sm}
\ee
The value of $a^d_{\rm sl}$ depends on $\Delta M_d, \Delta\Gamma_d$
and $\phi_d$, the parameters in the $B_d$ sector anologous to
those in Eq.~(\ref{asl-def}).
These parameters depend on the NP in the $B_d$ sector, which is
independent of the NP parameters in the $B_s$ sector that we 
are considering.
We therefore do not consider the measured values of $a^d_{\rm sl}$
as a direct constraint, but express it in terms of $\Delta M_d$, 
$\Delta\Gamma_d$, and $\phi_d$, whose experimental values are taken 
as inputs.

In the SM, we have $\phi_s(\rm SM) = 0.0041 \pm 0.0007$ \cite{lenz-nierste}. 
Note that if the dominating contribution to $\Gamma_{12s}$ were
from a pair of intermediate $c$ quarks, $\phi_s(\rm SM)$ would have been
equal to $-2\beta_s^{J/\psi \phi}$. Since the intermediate $u-c$ and
$u-u$ quark states give comparable contributions to $\Gamma_{12s}$,  
we have $\phi_s(\rm SM) 
\neq - 2 \beta_s^{J/\psi \phi}(\rm SM)$ \cite{lenz3802}.

\section{The statistical analysis}
\label{fit}

We perform a $\chi^2$ fit to the observed quantities $\Delta M_s, 
\Delta\Gamma_s, \beta_s^{J/\psi \phi}$ and $a^s_{\rm sl}$, 
using the NP parameters 
$|\m12np|, \, {\rm Arg}(\m12np), \, |\g12np|$ and ${\rm Arg}(\g12np)$.
We assume all the measurements to be independent for simplicity,
though the measurements of $\Delta\Gamma_s$ and $\beta_s^{J/\psi\phi}$
are somewhat correlated. 
The values of all the observables and their SM values are as given
in Sec.~\ref{measurements}.
In order to express them in terms of
$M_{12}, M_{12}^{\rm SM}, \Gamma_{12}$ and $\Gamma_{12}^{\rm SM}$,
one has to use Eq.~(\ref{dgs-def}) in addition.
In order to take into account the errors on the SM parameters,
we add the theoretical and experimental errors on our observed
quantities in quadrature.
 
Note that since we have four observable quantities and four parameters,
it is not surprising that we obtain the global minimum value of $\chi^2$
as $\chi^2_{min} =0$ when all the NP parameters are allowed to vary. 
The questions we address here are 
(i) what the preferred values of the NP parameters are, and 
(ii) to what confidence level (C.L.) a given set of NP parameters 
(or SM, which is a special case of NP with $\m12np = \g12np=0$)
is allowed.
The latter is obtained assuming all errors to be Gaussian.
Here we give our results in terms of the goodness-of-fit contours
for the joint estimations of two parameters at a time.
The $(1\sigma, 2\sigma, 3\sigma, 4\sigma)$ contours, that
are equivalent to $p$-values of $(0.3173, 0.0455, 0.0027, 0.0001)$, 
or confidence levels of $(68.27\%, 95.45\%, 99.73\%, 99.99 \%)$,
correspond to $\chi^2 = (2.295, 6.18, 11.83, 19.35)$, 
respectively.

\begin{figure}[h!]
\epsfig{file=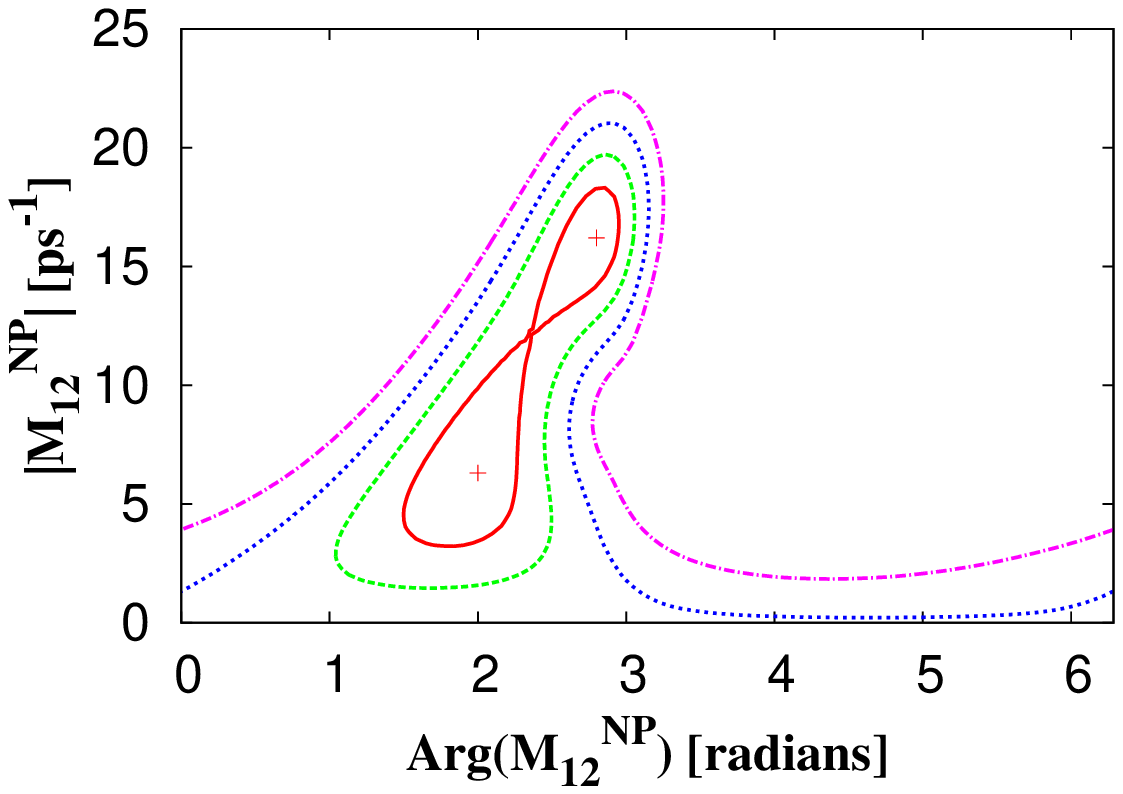,width=0.48\textwidth}
\caption{The $1\sigma$ (red/solid), $2\sigma$ (green/dashed),
$3\sigma$ (blue/dotted) and $4\sigma$ (pink/dot-dashed) goodness-of-fit contours  
in the $|\m12np| - {\rm Arg}(\m12np)$ plane, 
where the other NP parameters are marginalized over. The best-fit points, with $\chi^2=0$, are denoted 
by crosses. 
\label{fig12}}
\end{figure}

\begin{figure}[b]
\epsfig{file=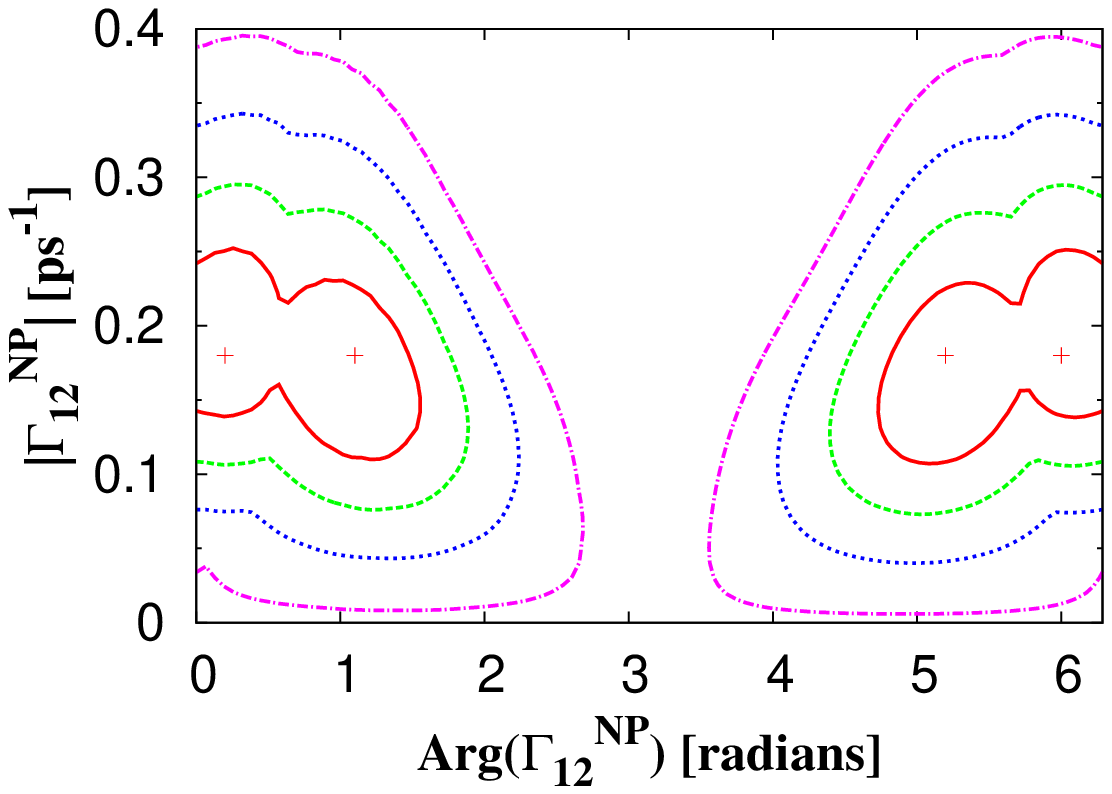,width=0.48\textwidth}
\caption{The $1\sigma$ (red/solid), $2\sigma$ (green/dashed),
$3\sigma$ (blue/dotted) and $4\sigma$ (pink/dot-dashed) goodness-of-fit contours  
in the $|\g12np| - {\rm Arg}(\g12np)$ plane, 
where the other NP parameters are marginalized over. The best-fit points, with $\chi^2=0$, are denoted 
by crosses.  
\label{fig34}}
\end{figure}

In Fig.~\ref{fig12}, we show the $1\sigma,2\sigma,3\sigma,4\sigma$ contours 
in the $|M_{12}| - {\rm Arg}(M_{12})$ plane, where the other NP
parameters are marginalized over. 
Clearly, we see a preference towards nonzero $|M_{12}^{NP}|$ as well
as nonzero ${\rm Arg}(\m12np)$ values. 
There are two best-fit points with $\chi^2=0$,
one at $\m12np \approx 6.3\exp(2.0\ i)$ ps$^{-1}$ and the other at
$\m12np \approx 16.2\exp(2.8\ i)$ ps$^{-1}$, shown with crosses
in Fig.~\ref{fig12}. Actually, each of these crosses is a superimposed
double, with two values of $\g12np$, as shown in Fig.~\ref{fig34}.
The points correspond to the constructive and destructive interference
between the SM and NP amplitudes in order to give the 
measured central values of $\Delta M_s$.
The region with $\m12np=0$, i.e. the $x$-axis, is outside
the $2\sigma$ region, indicating that it will be rather difficult
to fit the current data without some NP contribution to the
dispersive part of the $\bsbsbar$ mixing.
The contours also imply that $|\m12np| \lesssim 21.1$ ps$^{-1}$
to $3\sigma$.

In Fig.~\ref{fig34}, we show the goodness-of-fit contours 
in the $|\Gamma_{12}| - {\rm Arg}(\Gamma_{12})$ plane, marginalizing
over other two NP parameters. 
As the measurements do not determine the 
sign of $\Delta\Gamma_s$, for any particular value of $|\Delta\Gamma_s|$,
we perform the $\chi^2$ fit for both positive and negative
values, and keep the minimum $\chi^2$ of the two. This
doubles the number of best-fit solutions, and the two
best-fit points of Fig.~\ref{fig12} now split into four.
For $|\m12np|=6.3$, the solutions are $\g12np=
0.18\exp(6.0\ i)$ or $0.18\exp(5.2\ i)$, and for $|\m12np|
=16.2$, the corresponding solutions are $\g12np=
0.18\exp(0.2\ i)$ or $0.18\exp(1.1\ i)$ (both $\m12np$ and
$\g12np$ are in ps$^{-1}$, here, and also later where not
mentioned explicitly). Note that there is
a reflection symmetry about ${\rm Arg}(\g12np) = \pi$.
Again, a preference for nonzero values of 
$|\Gamma_{12}^{\rm NP}|$ is indicated, 
though ${\rm Arg}(\g12np)$ may vanish. 
The region with $\g12np=0$, i.e. the $x$-axis, is outside
the $4\sigma$ allowed region, indicating that NP contribution to
the absorptive part of the effective Hamiltonian is highly favored.
The contours also imply that $|\g12np| \lesssim 0.35$ at $3\sigma$.

\begin{figure}[h]
\epsfig{file=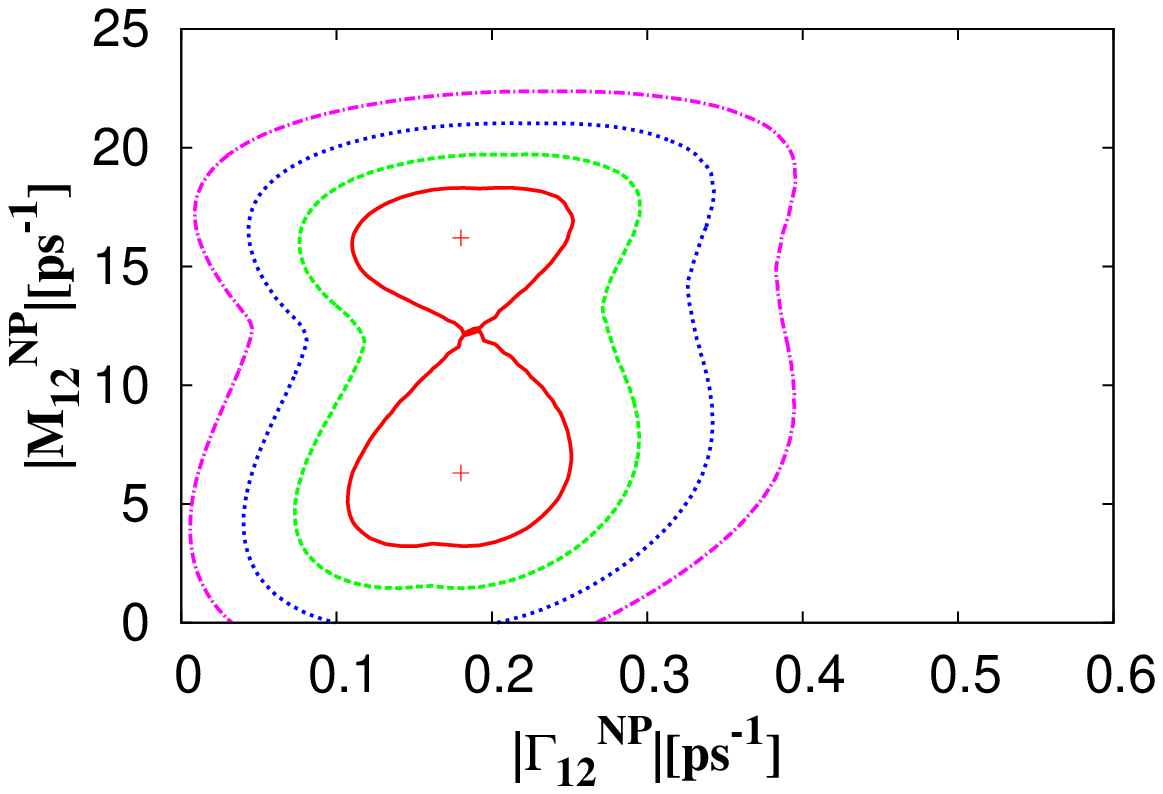,width=0.48\textwidth}
\caption{The $1\sigma$ (red/solid), $2\sigma$ (green/dashed),
$3\sigma$ (blue/dotted) and $4\sigma$ (pink/dot-dashed) goodness-of-fit contours  
in the $|\m12np| - |\g12np|$ plane, 
where the other NP parameters are marginalized over. The best-fit points, with $\chi^2=0$, are denoted 
by crosses. 
\label{fig13}}
\end{figure}

Fig.~\ref{fig13} displays the contours 
in the $|\m12np| - |\g12np|$ plane, and the two
NP phases are marginalized over. 
Not only does it show a preference for nonzero values of 
$\m12np$ and $\g12np$, but the $\m12np=0$ axis is
outside the $2\sigma$ allowed region and 
the $\g12np=0$ axis is outside the $4\sigma$ allowed region.
The best fit points are again superimposed doubles, whose
values can be read off from the discussion above.
The origin in this figure is the SM, which has
$\chi^2_{\rm SM}= 25.85$, and lies even outside the $4\sigma$
allowed region.
This dramatically quantifies the failure of the SM to accommodate
the current data. The reason is evident from eqs. (\ref{betas-value})
and (\ref{asl-def}); while $B_s\to J/\psi\phi$ prefers $\beta_s^
{J/\psi\phi}$ close to $\pi/8$ or $3\pi/8$, with a probability
minimum near $\beta_s^{J/\psi\phi}\approx \pi/4$, 
the measurement of $A^b_{\rm sl}$, and hence that of $a^s_{\rm sl}$,
prefers large $\tan\phi_s$, forcing $\beta_s^{J/\psi\phi}$ close to $\pi/4$.
This creates the tension between these two measurements.

\begin{figure}[h!]
\epsfig{file=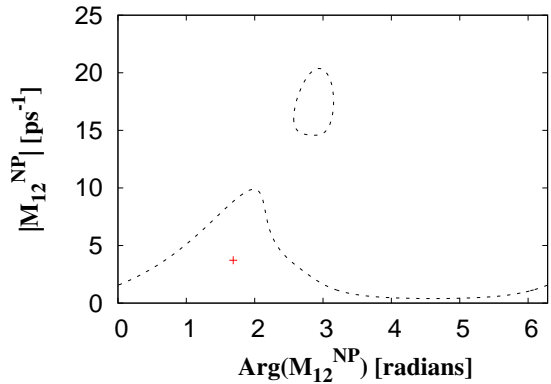,width=0.45\textwidth}
\caption{The $5\sigma$ goodness-of-fit contour in the 
$|\m12np| - {\rm Arg}(\m12np)$ plane, when $\g12np=0$, 
i.e. NP does not contribute to the absorptive part
of the effective Hamiltonian.
There are no points that are allowed to within $4\sigma$. The best-fit point, with $\chi^2=20.75$, is denoted 
by a cross. 
\label{fig120}}
\end{figure}

Fig.~\ref{fig13} also tells us that the models for which $\g12np=0$,
like R-parity conserving supersymmetry, universal extra dimension,
and extra scalars, fermions, or gauge bosons, cannot bring the 
tension down even to the $4\sigma$ range, unless the data moves
towards the SM expectations (and unless the new bosons are
flavor-changing so as to generate a nonzero $\g12np$). 
The best fit point with $\g12np=0$ 
has $\chi^2=20.75$ and corresponds to $\m12np = 3.72\exp(1.68\ i)$. 
This is further emphasized in fig.~\ref{fig120}, which shows the $5\sigma$
($p$ value of $10^{-6}$, $\chi^2= 27$) 
contour for those NP models where $\g12np$ is set to
vanish (within the closed contour above and under the open
contour below).

One may question the optimistic SM uncertainty for $\dgs$ as quoted in 
eq.\ (\ref{dgs-sm}). However, this has an almost negligible effect.
For example, if we increase the uncertainty by 50\%, neither the best 
fit points nor the confidence levels change significantly. The best fit 
point in fig.~\ref{fig120} has a $\chi^2$ minimum of 20.58 instead of 20.75.
The reason is the large deviation of $a^s_{\rm sl}$ from its SM value, 
to explain which we need a significant enhancement in $\tan\phi_s$.

\section{Preferred NP models}
\label{preferred}

From the results and discussion in the previous section,
it appears that:\\
(i) The SM by itself is strongly disfavored. Either $\m12np$ 
or $\g12np$ should be nonzero. \\
(ii) $\m12np\not= 0$ but $\g12np=0$ is also not allowed 
at $4\sigma$,
but the fit is marginally better than the SM. \\
(iii) The hypothetical case where $\g12np\not=0$ but $\m12np=0$
is also disfavored to more than $2\sigma$.
(This is a rather natural condition, since any interaction that
contributes to $\g12np$ will necessarily contribute to~$\m12np$.)

Most of the NP models can contribute significantly to $\m12np$.
Leading examples are the MFV models like minimal supersymmetry, 
universal extra dimensions, little-Higgs with T-parity, etc. Non-MFV
models like a fourth chiral generation, supersymmetry with R-parity
violation, two-Higgs doublet models, models with extra $Z'$, etc.\ can
also contribute significantly to $\m12np$.  

The NP models that can contribute significantly to $\g12np$, however,
are rather rare. This is because the NP contribution to the
absorptive part needs light particles in the final state, and
there are strong limits on the decays of $B_s$ to most of the possible
light final state particles. 
One of the few exceptions is the mode $\tau^+ \tau^-$, 
on which there is no available bound at this moment.
Thus, the NP that contributes to $\g12np$ has to do so via the
interaction $b \to s \tau^+ \tau^-$, but without affecting related
decays like $b \to s e^+ e^-$ or $b \to s \mu^+ \mu^-$ \cite{bauer}.
This can be achieved only in a limited subset of models, 
for example those with second and third generation scalar leptoquarks, 
R-parity violating supersymmetry \cite{dighe1},
or extra $Z'$ bosons \cite{alok-london}.
It turns out that the former can provide enough contribution to 
$\g12np$ to increase $\Delta\Gamma_s$ up to its current
experimental upper bound \cite{dighe1,dighe2}.
The amount of NP required for this is consistent with
the difference between the decay widths of $B_d$ and $B_s$
mesons ($\Gamma_s/\Gamma_d - 1 = (3.6\pm 1.8)\%$ \cite{hfag}),
and the recent  measurement of the branching
ratio of $B^+ \to K^+ \tau^+ \tau^-$, which is less than 
$3.3\times 10^{-3}$ at 90\% C.L.\ \cite{karim}.

One should note here that 
if the D\O~results on the dimuon charge asymmetry survive the 
test of time, it will be a clear indication of the presence of
a nonzero $\Gamma_{12s}^{\rm NP}$.
Such models are also favoured from the CDF and D\O~combined result 
on the allowed contours for $\beta_s^{J/\psi\phi}$ and $\Gamma_s$,
but we need to wait for the final Tevatron average.

\section{CPT Violation: the formalism}
\label{cptv-formalism}

The analysis till now is valid only if we assume CPT-invariance.
However, the CPT symmetry may be violated in theories that break
Lorentz invariance \cite{coleman-glashow}.
Indeed for local field theories, CPT violation requires Lorentz
violation \cite{greenberg1}. (This need not be true for nonlocal
field theories as well as for theories with noncommutative
space-time geometry, see \cite{chaichian}.) 
In general, CPT violation should result in differences in masses
and decay widths between particle-antiparticles pairs. 
However it may be easier to identify even through oscillation
experiments, which typically are sensitive to an interference
between the CPT-conserving and CPT-violating interactions.

While CPT violation in the $K$ system is severely constrained
through the mass difference between the neutral kaons \cite{pdg},
the bounds on the CPT violating parameters in the $B_d$ and $B_s$ systems 
are rather weak. In fact, the bounds for the $B_d$ sector are about
three orders of magnitude weaker than those for the $K$ sector
\cite{russell}.
The bounds on Lorentz-violating parameters using the data on B mesons 
can be found in \cite{kostel-bs} and references therein. 
Here we use a model-independent parameterization, like the one
earlier followed in \cite{datta} and recently used by two of us
\cite{sunando1}, and determine the preferred
parameter space using the data on $\bsbsbar$ oscillations.
Unlike \cite{kostel-bs}, we take both $A^b_{\rm sl}$ and $\beta_s^
{J/\psi\phi}$ data into account.

One should note that as a new physics option, CPT violation
is not exactly at the same footing as the models mentioned before.
However in the language of the effective Hamiltonian ${\cal H}$,
the CPT violation manifests itself naturally through in a difference 
between the diagonal elements of ${\cal H}$.
It is therefore interesting to see if  the constraints on the NP
coming from $\dms$ and $\dgs$ can be relaxed at all with these
additional degrees of freedom.
A posteriori, we will justify the discussion on CPT violation by 
showing that if the new physics indeed turns out to be without 
an absorptive part, CPT violation might help to explain the $\bsbsbar$ 
mixing data, albeit only marginally.

The CPT violation manifests itself in the effective Hamiltonian 
through the difference in the diagonal elements. 
We write the effective Hamiltonian in eq.\ (\ref{eff-H}) as
\be
\mathcal H = \begin{pmatrix}
 M_0 - \frac{i}{2} \Gamma_0 -\delta' & M_{12} - \frac{i}{2} \Gamma_{12} \\
 M_{12}^* - \frac{i}{2} \Gamma_{12}^* & M_0 - \frac{i}{2} \Gamma_0 +\delta'
 \end{pmatrix}  \; ,
\label{H-with-cpt} 
\ee 
and define the dimensionless CPT-violating complex parameter $\delta$ as
\be
\delta \equiv 
\frac{H_{22}-H_{11}}{\sqrt{{H}_{12} {H}_{21}}} 
=\frac{2\delta'}{\sqrt{{H}_{12} {H}_{21}}}\; ,
\ee
where $H_{ij} \equiv M_{ij} - \frac{i}{2} \Gamma_{ij}$. 

The eigenvalues of ${\cal H}$ are
\be
\lambda = \left(M_0 - \frac{i}{2}\Gamma_0\right) \pm \alpha y H_{12} \; ,
\label{evalues}
\ee
where 
$\alpha \equiv \sqrt{ H_{21} / H_{12} }$ and
$y \equiv \sqrt{ 1 + \delta^2/4 \, }$.
The corresponding mass eigenstates are
\bea
\left|B_{sH}\right> &= & p_1\left|B_s\right> + q_1\left|\overline{B}_s\right>
\; , \nonumber \\
 \left|B_{sL}\right> &= & p_2\left|B_s\right> - q_2\left|\overline{B}_s\right>
\; ,
\eea
with $|p_1|^2 + |q_1|^2 = |p_2|^2 + |q_2|^2 = 1$, and
\bea
\eta_1 & \equiv & \frac{q_1}{p_1} = 
\sqrt{\frac{H_{21}}{H_{12}}} 
\left( \sqrt{1 + \frac{\delta^2}{4}} + \frac{\delta}{2}\right) \,,\nonumber\\
\eta_2 & \equiv & \frac{q_2}{p_2} = 
\sqrt{\frac{H_{21}}{H_{12}}} 
\left( \sqrt{1 + \frac{\delta^2}{4}} - \frac{\delta}{2}\right) 
\,.
\eea
Clearly, CPT invariance corresponds to $\eta_1 = \eta_2$.

Let us now determine the dependence of our four observables on
the CPT-violating parameters.
The differences in masses and decay widths of the eigenstates  are
related to the difference in eigenvalues as
\be
\lambda_1-\lambda_2 = \Delta M + \frac{i}{2}\Delta \Gamma\,,
\ee
where $\lambda_1$ and $\lambda_2$ are ordered such that 
${\rm Re}(\lambda_1 - \lambda_2) >0$.
From Eq.~(\ref{evalues}), 
\bea
\Delta M &=& M_1 - M_2 = 2 {\rm Re} ( \alpha y H_{12}) \; , 
\label{dm-cpt} \\  
\Delta\Gamma &=& \Gamma_2 - \Gamma_1 = 4 {\rm Im} (\alpha y H_{12}) 
\label{dgs-cpt} \; .
\eea
Since $|\Gamma_{12}| \ll |M_{12}|$, we can write
\bea
\alpha H_{12} & = &  
|M_{12}| \left[ 1 -\frac{1}{4} \frac{|\Gamma_{12}|^2}{|M_{12}|^2} 
- i {\rm Re}\left(\frac{\Gamma_{12}}{M_{12}}\right)\right]^{\frac{1}{2}} 
\nonumber \\
& \approx & |M_{12}| \left[ 1 - \frac{i}{2} 
{\rm Re}\left(\frac{\Gamma_{12}}{M_{12}}\right)\right] \,.
\eea
Then Eqs.~(\ref{dm-cpt}) and (\ref{dgs-cpt}) yield 
\bea
\Delta M & \approx & |M_{12}| 
\left[2 {\rm Re}(y) + {\rm Im}(y) 
{\rm Re}\left(\frac{\Gamma_{12}}{M_{12}}\right) \right]\; , 
\label{dm-cpt-approx} \\
\Delta\Gamma & \approx &  |M_{12}| \left[  4 {\rm Im}(y) 
- 2 {\rm Re}(y) {\rm Re}\left(\frac{\Gamma_{12}}{M_{12}}\right) \right] \; .
\label{dgs-cpt-approx}
\eea
The dependence on the CPT-violating parameter $\delta$ appears
entirely through $y$. 

Let us pause here for a moment and find what the above two
equations tell us about the allowed parameter space.
Let us first focus on the best constraint, $\Delta M_s$, and work
in the limit where $\Gamma_{12}/M_{12}$ is negligible. $|M_{12}|$,
and hence $\m12np$, can be arbitrarily large, as Re($y$) can be
made arbitrarily small by an appropriate choice of $\delta$. 
Similarly, Re($y$)
can be quite large (albeit compatible with other constraints) as long
as there is a near-perfect cancellation between the SM and NP
mixing amplitudes, making $|M_{12}|$ small.
However, the smallness of $\Delta\Gamma/\Delta M$ constrains
${\rm Im}(y)/{\rm Re}(y)$ to be small, thus indicating that
$y$ is almost real. 
Since $y = \sqrt{1 + \delta^2/4}$, this implies that
$\delta^2$ is almost real and ${\rm Re}(\delta^2) \gtrsim -4$.
Therefore, one would expect that $\delta$ is either almost real,
or it is almost imaginary, but with $|{\rm Im}(\delta)| <2$.

Now let us consider the CP-violating observables $\beta_s^{J/\psi\phi}$
and $a^s_{\rm sl}$.
The effective value of the former may be obtained in the presence
of CPT violation by considering the decay rates of $B_s$ and 
$\overline{B}_s$ to a final CP eigenstate $f_{CP}$ as 
\cite{sunando1}:
\bea
\Gamma(B_s(t) \to f_{CP}) & = & |A_f|^2 \big[ 
|f_+(t)|^2  + |\xi_{f_1}|^2 |f_-(t)|^2 +  \nonumber \\
 & & \phantom{space}  2 {\rm Re}( \xi_{f_1} f_-(t) f_+^*(t)) \big] \; , 
\label{evol-B} \\
\Gamma(\overline{B}_s(t) \to f_{CP}) & = & 
|\frac{A_f}{\eta_2}|^2 
\big[ |f_-(t)|^2  + |\xi_{f_2}|^2 |\overline{f}_+(t)|^2 +  \nonumber \\
 & & \phantom{space}  2 {\rm Re}( \xi_{f_2} \overline{f}_+(t) f_-^*(t)) 
\big] \; ,
\label{evol-Bbar}
\eea
with
\be
\xi_{f_1} \equiv \eta_1 \frac{\overline{A}_{{f}}}{A_f} \; , \quad
\xi_{f_2} \equiv \eta_2 \frac{\overline{A}_{{f}}}{A_f} \; , \quad
\omega \equiv \frac{\eta_1}{\eta_2} \; .
\ee
Here $A_f$ and $\overline{A}_f$ are the amplitudes for the processes 
$B_s \to f_{CP}$ and $\overline{B}_s \to f_{CP}$, respectively. 
The time evolutions are given by
\bea
f_-(t) & = & \frac{1}{1 + \omega} (e^{-i \lambda_1 t} - e^{-i \lambda_2 t})
\; , \nonumber \\
f_+(t) & = & \frac{1}{1 + \omega} (e^{-i \lambda_1 t} + \omega e^{-i \lambda_2 t})
\; ,\nonumber \\
\bar{f}_+(t) & = & \frac{1}{1 + \omega} (w e^{-i \lambda_1 t} + e^{-i \lambda_2 t})
\; .\label{f-defs}
\eea
The final state in $B_s \to J/\psi \phi$ is not a CP eigenstate, but 
a combination of CP-even and CP-odd final states, which may be 
separated using angular distributions.
With the transversity angle distribution \cite{transversity-angle}, 
the time-dependent decay rate to the CP-even state is given by
the coefficient of $(1+\cos^2 \theta)$, while the time-dependent
decay rate to the CP-odd state is given by the coefficient of
$\sin^2\theta$.

The value of effective $\beta_s^{J/\psi\phi}$ in this process is determined by
writing the time evolutions (\ref{evol-B}) and (\ref{evol-Bbar}) in
the form
\bea
\Gamma(B_s(t) \to f_{CP}) & = & 
c_1 \cosh(\Delta\Gamma_s t/2) + 
c_2 \sinh(\Delta\Gamma_s t/2) + \nonumber \\
& & c_3 \cos(\Delta M_st) + c_4 \sin(\Delta M_st) \; , \\
\Gamma(\overline{B}_s(t) \to f_{CP}) & = & 
\bar{c}_1 \cosh(\Delta\Gamma_s t/2) + 
\bar{c}_2 \sinh(\Delta\Gamma_s t/2) + \nonumber \\
& & \bar{c}_3 \cos(\Delta M_st) + 
\bar{c}_4 \sin(\Delta M_st) \; . 
\label{trig-form}
\eea
The direct CP violation in $B_s \to J/\psi \phi$ is negligible;
i.e. $|\overline{A}_f/A_f|\approx 1$. 
Also, $|\Gamma_{12}/M_{12}| \ll 1$, so that in the absence of
CPT violation, $|\eta_1| = |\eta_2| =1$.
Then in terms of $\xi_f \equiv \xi_{f_1} = \xi_{f_2} = 
\alpha \overline{A}_f / A_f$, one can write 
\be
\frac{c_4}{c_1} = - \frac{\bar{c}_4}{\bar{c}_1}
= \frac{2 {\rm Im}(\xi_f)}{1 + |\xi_f|^2} 
\approx - \eta_{CP} \sin(2\beta_s^{J/\psi\phi}) \; ,
\ee
where $\eta_{CP}$ is the CP eigenvalue of $f_{CP}$. 

When CPT is violated, the effective phases $\beta_s^{J/\psi\phi}$ 
and $\bar{\beta}_s^{J/\psi\phi}$ measured through 
$B_s(t)$ and $\overline{B}_s(t)$ decays, respectively,
will turn out to be different.  
Indeed, the difference between these effective phases will be
a clean signal of CPT violation.
\begin{widetext}
\bea
\sin(2\beta_s^{J/\psi\phi}) & = &
-\eta_{CP} \frac{2 [-{\rm Im}(\omega) - {\rm Re}(\xi_{f_1}) {\rm Im}(\omega)
+ {\rm Im}(\xi_{f_1}) + {\rm Im}(\xi_{f_1}) {\rm Re}(\omega)]}
{[1 + |\omega|^2 + 2 |\xi_{f_1}|^2 + 2 {\rm Re}(\xi_{f_1})
- 2 {\rm Re}(\xi_{f_1}) {\rm Re}(\omega) 
- 2 {\rm Im}(\xi_{f_1}) {\rm Im}(\omega)]} \; , 
\label{betaeff-B} \\
\sin(2 \bar{\beta}_s^{J/\psi\phi}) & = &
-\eta_{CP} \frac{2 [- |\xi_{f_2}|^2 {\rm Im}(\omega) + {\rm Re}(\xi_{f_2}) {\rm Im}(\omega)
+ {\rm Im}(\xi_{f_2}) + {\rm Im}(\xi_{f_2}) {\rm Re}(\omega)]}
{[2 + |\xi_{f_2}|^2 (1 + |\omega|^2) - 2 {\rm Re}(\xi_{f_2})
+ 2 {\rm Re}(\xi_{f_2}) {\rm Re}(\omega) 
- 2 {\rm Im}(\xi_{f_2}) {\rm Im}(\omega)]} \; ,
\label{betaeff-Bbar}
\eea
\end{widetext}
Though the analysis of the $B_s$ and $\overline{B}_s$ modes needs to
be performed separately, here we assume identical detection and
tagging efficiencies for both, and use  
the average of Eq.\ (\ref{betaeff-B}) and Eq.\ (\ref{betaeff-Bbar}) for our fit.

The semileptonic CP asymmetry $a^s_{\rm sl}$ is measured through the
``wrong-sign'' lepton signal:
\be
a^s_{\rm sl} = \frac{\Gamma(\overline{B}_s(t)\rightarrow \mu^+ X) -
              \Gamma({B}_s(t)\rightarrow \mu^- X)}
             {\Gamma(\overline{B}_s(t)\rightarrow \mu^+ X) +
              \Gamma({B}_s(t)\rightarrow \mu^- X)}\; .
              \label{aslq1}
\ee
Here,
\bea
\Gamma({B}_s(t)\rightarrow \mu^- X) & = &  |\eta_1 f_- A(B_s \to \mu^+ X)|^2 \; , \\
\Gamma(\overline{B}_s(t)\rightarrow \mu^+ X) & = &  
|(f_-/\eta_2)  A(\overline{B}_s \to \mu^+ X)|^2 \; , 
\eea
and since  
$|A(\overline{B}_s \to \mu^+ X)|=  |A(B_s \to \mu^+ X)|$,
\be
a^s_{\rm sl} 
= \frac{\frac{1}{|\eta_2|^2} - |\eta_1|^2}{\frac{1}{|\eta_2|^2} + |\eta_1|^2} 
= \frac{1 - |\alpha|^4}{1 + |\alpha|^4}\; ,
\label{asl-cpt}
\ee
which is independent of the CPT-violating parameter $\delta$. 
That the semileptonic asymmetry does not contain a CPT violating term in the
leading order was also noted earlier \cite{pais-treiman-shalom}.

\section{CPT violation: The statistical analysis}
\label{cptv-fit}

\begin{figure}[h!]
\epsfig{file=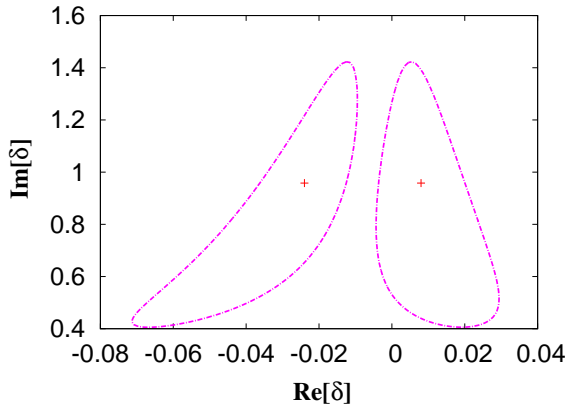,width=0.45\textwidth}
\caption{The $4\sigma$ goodness-of-fit contours 
in the ${\rm Re}(\delta) - {\rm Im}(\delta)$ plane, when 
the only relevant NP contribution is CPT violating, parameterized
entirely by $\delta$.
There are no points that are allowed to within $3\sigma$. 
The crosses show the best fit points, with $\chi^2=16.4$.
\label{fig-cpt-only}}
\end{figure}

In this Section, we perform a $\chi^2$-fit to the observables
$\Delta M_s$, $\Delta\Gamma_s$, the effective phase $\beta_s^
{J/\psi\phi}$, and $a^s_{\rm sl}$.
Let us first assume that there is no CPT-conserving NP
contribution coming from $\m12np$ and $\g12np$, so that the
only relevant NP contribution is CPT violating, and is
parameterized by ${\rm Re}(\delta)$ and ${\rm Im}(\delta)$.
The allowed parameter space is shown in Fig.~\ref{fig-cpt-only}.
It turns out that in this case, the value of $\chi^2_{\rm min}$ is 
$\approx$ 16.4 
(at $\delta = 0.008 + 0.958 \ i$ and $\delta = -0.024 + 0.958 \ i$),
marginally better than the one obtained 
in the $(\g12np=0, \m12np \neq 0)$ case discussed above in
Fig.~\ref{fig120}. There
are some, albeit small, regions in the parameter space
that are allowed to $4\sigma$. 
However a fit good to $3\sigma$ or better is still not possible.

We therefore need to add the CPT-conserving NP to the CPT-violating
contribution.
However we have already seen in the preceding section that 
$\m12np$ and $\g12np$ together are capable of explaining the
data by themselves. 
Therefore the fit using $\delta$, $\m12np$ as well as $\g12np$
is redundant. With six
independent parameters and only four observables, not only is
$\chi^2_{min}=0$ guaranteed, but no effective limits on CPT-conserving
and CPT-violating parameters are generated.

We, therefore, go directly to the possibility where there is CPT-conserving
NP, but without an absorptive part: $\g12np=0$. We have already 
observed  (Fig.~\ref{fig120}) that the entire region in the 
$|\m12np|-{\rm Arg}(\m12np)$ is outside the $4\sigma$ region 
in such a scenario. We would now ask
what happens if we enhance the two-parameter NP with two more
CPT violating parameters, {\em viz.}, Re($\delta$) and Im($\delta$).
This scenario is
interesting because, as we have seen before, only very specific
kind of NP can contribute to $\g12np$, which would
be tested severely in near future. In case no evidence for
the relevant NP is found (e.g. the branching ratio of 
$B_s \to \tau^+ \tau^-$ is observed to be the same as its SM
prediction), the next step would be to check if CPT violation,
along with the NP contribution through $\m12np$,
would be able to account for the anomalies. 
For example, one
may want to determine $\beta_s$ and $\bar\beta_s$ of
Eqs.~(\ref{betaeff-B}) and (\ref{betaeff-Bbar}) separately
and see whether they are different.

\begin{figure}[h!]
\epsfig{file=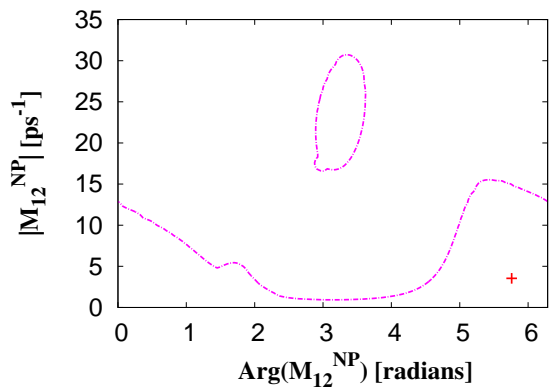,width=0.45\textwidth}
\caption{The $4\sigma$  goodness-of-fit 
contours in the $|\m12np| - {\rm Arg}(\m12np)$ plane, when $\g12np=0$, 
i.e. NP does not contribute to the absorptive part
of the effective Hamiltonian.
The CPT-violating complex parameter $\delta$ has been
marginalized over. 
There are no points that are allowed to within $3\sigma$.
The cross shows the best fit point, with $\chi^2=14.3$.
\label{m12-phase-0}}
\end{figure}

Fig.~\ref{m12-phase-0} shows the situation in the 
$|\m12np|-{\rm Arg}(\m12np)$ plane.
As compared to Fig.~\ref{fig120}, one can see that once we marginalize 
over $\delta$, we now have some regions allowed to within $4\sigma$
(within the closed contour above and below the open contour),
but none within $3\sigma$.
Indeed, $\chi^2_{min} = 14.3$ at $\m12np = 3.54 \exp(5.76\ i)$.
This clearly does not improve the goodness-of-fit substantially,
indicating that there is no good alternative for $\g12np$.

\begin{figure}[h!]
\epsfig{file=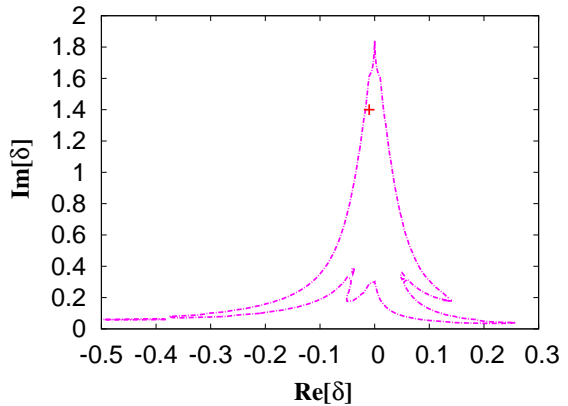,width=0.45\textwidth}
\caption{The $4\sigma$  
goodness-of-fit 
contours in the ${\rm Re}(\delta) - {\rm Im}(\delta)$ plane, 
when the complex NP parameter $\m12np$ is marginalized over, while
$\g12np$ has been constrained to vanish.
There are no points that are allowed to within $3\sigma$.
The cross shows the best fit point, with $\chi^2=14.3$.
\label{red-rem-0}}
\end{figure}

Fig.~\ref{red-rem-0} shows the situation in the complex $\delta$
plane, when $\m12np$ has been marginalized over. 
The best-fit point corresponds to $\delta = -0.01 + 1.40 \ i$, which gives
$\chi^2_{min} = 14.3$ as mentioned earlier.
The CPT conserving point ($\delta=0$) lies outside the $4\sigma$ region.
As expected from the discussion in Sec.~\ref{cptv-formalism},
the allowed values of $\delta$ are close to the ${\rm Re}(\delta)$
or ${\rm Im}(\delta)$ axis, with $|{\rm Im}(\delta)|$
restricted to 2. One observes that the
current data allows rather large ($\sim 1$) positive values of 
${\rm Im}(\delta)$ at $4\sigma$.

\section{Conclusion}
\label{concl}

Any flavor-dependent new physics model can in general affect both mass and 
width differences in the $B_s$--$\overline{B}_s$ system. 
It can also affect the CP-violating phase, as well as the dimuon asymmetry, 
which was found by the D\O~ collaboration to have an anomalously large value. 
With these four observables, one can constrain the free parameters 
of the new physics model. 
We have used the model independent approach where we consider the 
effective $\bsbsbar$ mixing Hamiltonian ${\cal H}$ 
and parameterize the NP through its contribution to ${\cal H}$.
We quantify the goodness-of-fit for the SM and NP parameter values 
by performing a combined $\chi^2$-fit to all the four measurements.
The tension of the data with the SM is clear by the high value of
$\chi^2$ at the SM. 
Moreover, it is observed that we need NP to contribute to the dispersive 
as well as absorptive part of the off-diagonal elements of ${\cal H}$ 
in order for the current data to be explained.
The absorptive contribution, in particular, can be obtained from 
a very limited set of models, which will be severely tested in near future.

We also introduce the possibility of CPT violation by adding
unequal NP contributions to the diagonal elements of ${\cal H}$.
We explicitly show how CPT violation might affect the observables,
especially dwelling on the effect on $\beta_s^{J/\psi \phi}$.
Taken alone, the CPT violation cannot affect the dimuon asymmetry,
and it can make the fit to the $\bsbsbar$ mixing data
only marginally better.
In combination with a CPT conserving NP, it can enhance the
allowed parameter space for that NP, however it does not seem to be
able to obviate the need of an absorptive contribution from NP.

The data on all the observables considered in this paper is still 
relatively preliminary, the deviations from the SM are only at about
2-3$\sigma$ level, and future data may either confirm these deviations
or expose them as statistical fluctuations. 
If the errors and uncertainties shrink keeping the central
values more or less intact, this will mean:
\begin{itemize}

\item The SM is strongly disfavored.
Moreover, the relevant NP should be flavor-dependent, as we do not see 
much deviation in the $\bdbdbar$ sector.

\item The NP models that do not contribute to the absorptive amplitude of
the $\bsbsbar$ mixing are also strongly disfavored if CPT is conserved.
The best bets are those NP models that provide both dispersive
and absorptive amplitudes in the $\bsbsbar$ mixing. This also gives
rise to new decay channels for $B_s$. For example, one might find the
branching ratio of $B_s\to\tau^+\tau^-$ enhanced significantly from its SM
expectation.

\item Without any CPT-conserving NP, only CPT violation is only 
of marginal help, as it cannot enhance the semileptonic asymmetry.
Even in combination with the CPT-conserving dispersive NP, 
it cannot allow regions in the parameter space to better than $3\sigma$.

\end{itemize}

To summarize, the NP models that contribute an absorptive part
to $\bsbsbar$ mixing seem to be essential if one wants to explain
the data on $\beta_s^{J\psi \phi}$ and $A^b_{sl}$ simultaneously.
There is only a limited set of such models, and they will be 
severely tested in near future.
In the scenario that such an absorptive NP contribution is
ruled out, one may have to resort to CPT violation in order
to explain the data.
A prominent signature of such a CPT violation would be a difference
in $\beta_s^{J\psi /\phi}$ and $\bar{\beta}_s^{J/\psi\phi}$ as shown
in eqs.\ (\ref{betaeff-B}) and (\ref{betaeff-Bbar}).

\section*{Acknowledgements}

SKP acknowledges CSIR, Government of India, for a research fellowship. 
The work of AK was supported by CSIR, Government of India, 
and the DRS programme of the University Grants Commission. 


\end{document}